# THE 2016 AL-MISHRAQ SULPHUR PLANT FIRE: SOURCE AND RISK AREA ESTIMATION

Oscar Björnham[†], Håkan Grahn, Pontus von Schoenberg, Birgitta Liljedahl, Annica Waleij, Niklas Brännström

[†] Corresponding author:
Oscar Björnham
oscar.bjornham@foi.se

# 1. ABSTRACT

On October 20 2016, Daesh (Islamic State) set fire to the sulphur production site Al-Mishraq as the battle of Mosul became more intense. A huge plume of toxic sulphur dioxide and hydrogen sulphide caused comprising casualties. The intensity of the $SO_2$ release was reaching levels of minor volcanic eruptions which was observed by several satellites. By investigation of the measurement data from the MetOp-A, MetOp-B, Aura, and Meteosat-10 satellites we have estimated the time-dependent source term for sulphur dioxide with 92 kt $SO_2$ released into the atmosphere during six days. The long-range dispersion model PELLO was utilized to simulate the atmospheric transport over the Middle East. The ground-level concentrations predicted by the simulation were compared with observation from the Turkey National Air Quality Monitoring Network. Finally, the simulation data provided, using a probit analysis, an estimate of the risk area at ground level which was compared to reported urgent medical treatments.

# 2. INTRODUCTION

Scorched earth tactics are not a novel modus operandi in conflict situations. Although banned under Article 54 of Protocol I of the 1977 Geneva Conventions, it is regularly used as a warfare tactic, by both state actors and non-state actors. For instance, during the first Gulf War in 1991, retreating Iraqi forces set fire to over 600 oil wells as well as numerous oil filled low-lying areas, such as trenches. More recently, and in particular since the beginning of the Mosul offensive, Daesh (ISIS), have revived these tactics including deliberate oil fires and at least one attack on a chemical plant, namely the sulphur plant in Al-Mishraq which had one of largest sulphur deposits in the world (Zwijnenburg, 2016, Kalin, 2016, Al-Yaseen and Niles, 2016, Sis, 2016). This is however not the first time the Al-Mishraq plant is set on fire, in an alleged act of arson it burnt in June 2003. Then the fire burned for approximately four weeks causing a release of ~600 kt sulphur dioxide (Carn et al., 2004). The resulting toxic plume dispersed over a large area causing acute short term injuries in exposed military staff and population (Baird et al., 2012) and is possibly linked to long term adverse medical effects including constrictive bronchiolitis (USAPHC, 2012).

In the current situation, with the October advancement on Mosul, Daesh created a complex battle environment with the attack and fire at the Al-Mishraq sulphur plant combined with oil fires and the alleged use of chemical weapons (Deutsch, 2016). These acts resulted in an amplification of the already present humanitarian crises in the region. Moreover, humanitarian aid personnel and military deployed personnel were also affected thereof. This once again accentuates the need for accurate and timely health threat assessment in conflict areas.

In the ideal world, real-time air sampling and environmental monitoring would be conducted in theatre. This is however rarely the case due to the nature of conflict zones. At best sampling could be conducted in the immediate environment of deployed troops. Nevertheless, reliable data are needed to obtain sound information regarding levels of pollutants and, if necessary, to formulate appropriate protective measures. To better inform medical intelligence (Wikström et al., 2016) such data could be supplied through dispersion modeling, but this requires a reliable source term, i.e. an assessment of at which rate the pollutant is injected in the atmosphere. In this paper we show that remote sensing through satellite images of $SO_2$ can be utilized to provide a rapid source estimate for dispersion modeling. The dispersion model results are compared with air measurements from the Turkey National Air Quality Monitoring Network.

**Abbrevations**
| | |
|---|---|
| ISIS | Islamic State of Iraq and Syria |
| UTC | Coordinated Universal Time |
| MSE | Mean square error |
| lat | Latitude |
| long | Longitude |

## 3. SATELLITE DATA

In absence of a detailed in situ description of the fire at the Al-Mishraq sulphur mine, satellite images provide the best basic data of the release rate. Today there exists a number of different systems that measure the sulphur dioxide load in the atmosphere: for example NASA's Aura satellite carries the Ozone Monitoring Instrument (OMI) (Levelt et al., 2006), and EUMETSAT and ESA's satellites MetOp-A and MetOp-B each carries an instrument called the Global Ozone Monitoring Experiment-2 (GOME-2) (Hassinen et al., 2016). Both OMI and GOME-2 utilize hyperspectral imaging and study the solar backscatter radiation to detect aerosols and trace gases. Level 2 data (processed data) from the GOME-2 instruments are disseminated via the Satellite Application Facility on Ozone and Atmospheric Chemistry Monitoring project (O3M SAF). Amongst the level 2 products available we find $SO_2$ total vertical column data. The data is provided using data points in units of molecules/cm$^2$, where each point represents a surface area of 40x40 km$^2$ and 40x80 km$^2$ for MetOp-A and MetOp-B, respectively. Unfortunately there are factors that cause uncertainties in the $SO_2$ measurements such as cloud coverage and interference between $SO_2$ and ozone (Fioletov et al., 2013). The target uncertainty for GOME-2 $SO_2$ measurements is 50% (Hassinen et al., 2016). $SO_2$ can also be detected in the thermal infrared spectrum, for example by the Spin Enhanced Visible and Infrared Imager (SEVIRI) system located on the satellite Meteosat-10 and the Moderate Resolution Imaging Spectroradiometer (MODIS) instruments on the satellites Aqua and Terra. There are other satellite data, e.g. OMPS, IASI and AIRS, that can be used to derive information about the Al-Mishraq fire, that have not been considered in this study.

Combining MetOp-A and MetOp-B measurements gives a near complete coverage of the area of interest: the lat-long box 30°E - 60°E, 25°N - 45°N. There may occur overlaps of MetOp-A and MetOp-B observations which are resolved by using the mean value in these areas. However, since the satellites are not geostationary they only probe the area of interest a few minutes every day resulting in discrete snapshots rather than a continuous development. In contrast, the Meteosat-10 satellite is indeed geostationary and SEVIRI covers the region of interest providing images every 15 minutes. We use the SEVIRI images for qualitative assessments of the source, however not for quantitative estimates of $SO_2$ loadings. The independent datasets for GOME-A/B and OMI show good agreement which serves as a confirmation of the data quality. Given all these prerequisites, we chose to use the GOME-A/B measurements in the source estimation process.

The $SO_2$ loading of the atmosphere is given as Vertically Column Density Corrected (VCDC) values per data point in the GOME-A/B dataset. Processed data of this type is usually refered to as level 2 data. As the name suggest these VCDC values have been corrected for a number of physical and chemical phenomena that interfere with $SO_2$ remote measurements, including for example correction for interference between $SO_2$ and ozone (Fioletov et al., 2013). $SO_2$ VCDC values are given in Dobson units (DU), where 1 DU corresponds to $2.69 \times 10^{20}$ molecules per square meter when the column is integrated vertically. The translation from sensor data to the corresponding concentrations in DU depends on the actual height of the $SO_2$ plume in the given column. The GOME-A/B instruments cannot detect the height of the $SO_2$ plume, hence the level 2 data VCDC values are given for four different assumed heights of the plume (1.0 km, 2.5 km, 6.0 km and 15.0 km). We have implemented a method where the plume height is differentiated over the region and we therefore interpolate the VCDC value in each position to the local height of the $SO_2$ plume (where the latter is inferred from the dispersion model results). Further on, the dataset includes quality flags that indicate if data points may be subject to errors or increased uncertainties. All data points with any such indication have been removed in this analysis, i.e. we use QualityFlag = 0. Furthermore, we apply the following limitations to improve the quality of the data points used: IndexInScan ≤ 2 (only forward scans), ViewMode = 256 (only day measurements) and SolarZenithAngleCentre ≤ 75 (only appropriate angle for the sun). In this study only a small fraction of the provided points are excluded for any of these limitation leaving the majority of the data points available for the source term estimation.

Presence of clouds interferes with the measurements of $SO_2$ mainly due to the fact that the instrument is unable to detect the part of the plume that is located below the cloud. The affected column densities can be compensated for the presence of clouds by altering the air mass factor, this procedure will however implicitly add a ghost-column of $SO_2$ below the cloud representing the unavailable region for the satellite (Theys et al., 2015). The uncertainties will still be significant in cases where a large fraction of the $SO_2$ is located in or below the cloud and in particular if the vertical distribution of $SO_2$ is anisotropic. The datasets from the GOME-systems include measurements of the cloud cover. Fortunately there is negligible cloud cover over Al-Mishraq during the fire. However, there is a prevailing cloud cover in the northern part of the region of interest which interfere with the measurements as the plume drifts into this region mainly on October 25 - 27. A plume is still visible during these days but the uncertainties are expected to increase due to the clouds.

## 4. ESTIMATING THE SOURCE TERM

The GOME-2 datasets provide a measure of the total $SO_2$ load in the atmosphere. To yield the desired source term this $SO_2$ load has to be converted to a flux of $SO_2$ from the Al-Mishraq plant. This problem has been studied in conjunction with volcanic eruptions and Theys et al. present a survey of methods of how to derive the flux (Theys et al., 2013). To get an accurate estimate of the flux, the satellite data should be interpreted with respect to how $SO_2$ is dispersed in the atmosphere including any chemical reactions that it may undergo (Stohl et al., 2011). In our method (Grahn et al., 2015) we use the Lagrangian particle random displacement model PELLO (Lindqvist, 1999) to describe how the $SO_2$ is transported in the atmosphere using 2.5 million model particles. The benefits from this approach are manifold which we describe below. PELLO is fed by meteorological data from the European Centre for Medium-Range Weather Forecasts (ECMWF) with a time resolution of 3 hours. Removal of $SO_2$ by wet and dry deposition is taken into account in PELLO as well as a first order kinetics which depletes $SO_2$ where the e-folding time has been estimated to 48 hours due to the amount of cloud cover during this time period (Beirle et al., 2014). This value corresponds to a half-life time of 33 hours.

Using $SO_2$ satellite images from SEVIRI we conclude that $SO_2$ emissions started in the morning of October 20 and ceased in the early hours on October 27. From satellite visibility we infer that a significant release of SO$_2$ began at approximately 07:00 on October 20. All times given in UTC throughout this paper. Two SEVIRI images with a time difference of 5 hours are presented in Figure 1 and show the plume fluctuating over time. The location of the fire was captured before and after the event in high resolution in the visual spectrum by the satellite WorldView-3, showing that it was mainly open air sulphur stockpiles that were set ablaze (SATCEN, 2016). Furthermore the images confirm that the fire caused a significant reduction in the sulphur stockpiles and that an area of approximately 500 x 1500 m$^2$ burned during this event which is the horizontal extension we use for the source term in the simulations.

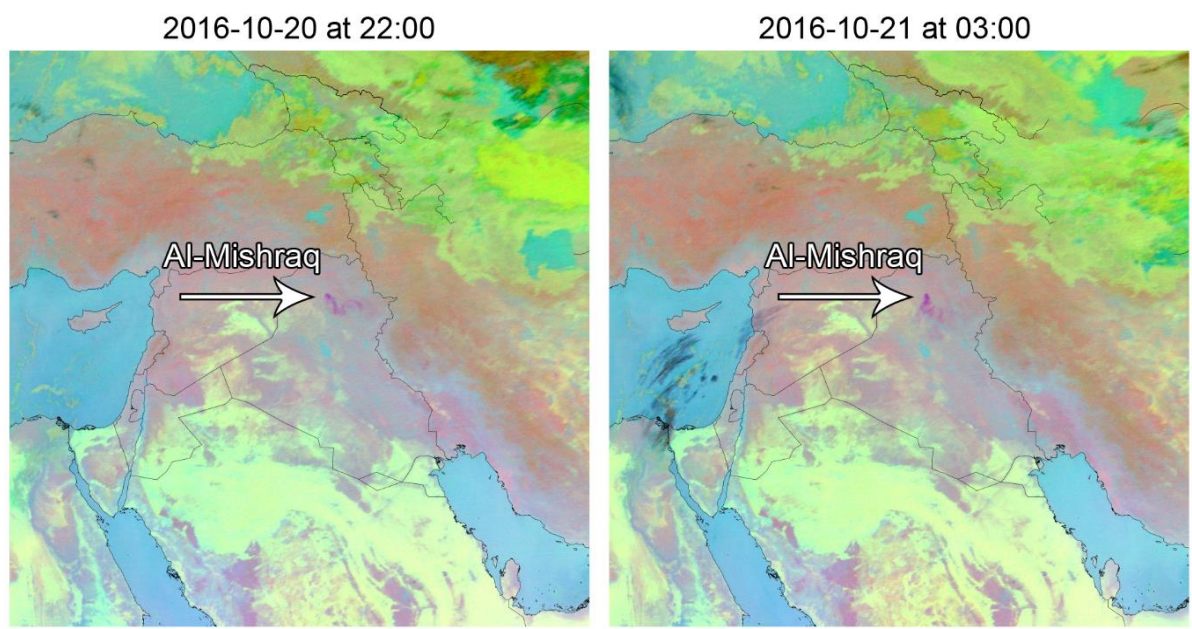

*Figure 1. Infrared image over the Middle East taken by SEVIRI taken at 22:00 October 20 2016 (panel a) and at 03:00 October 21 2016 (panel b). The purple plumes show the release of SO$_2$ at the initial stage of the fire which predominately spreads southward for the first days.*

### 4.1. SOURCE DESIGN

Since the MetOp-A and MetOp-B satellites both passed Iraq around 07:00 every morning we only have one, assembled, snap-shot of the SO$_2$ load each day. In lieu of more time-resolved information we apply the long-range dispersion model PELLO. The temporal resolution is increased by investigation of the development of the simulated plume since the dispersion model operates on a timescale of seconds. For the source term, we divide each day into time intervals of 6 hours. The time intervals for the virtual sources are chosen to harmonise with the satellite data setting daily break points at 01:00, 07:00, 13:00 and 19:00. The release rate is kept constant in each interval. By this temporal breakdown of the day we allow for a more precise estimate of the development of the fire process.

Photographs of the event, e.g. by Thaier Al-Sudani/Reuters, make us conclude that SO$_2$ was released from ground level to a considerable height, i.e., the heat caused a significant plume rise. The actual vertical distribution of the release of the SO$_2$ is unknown and is impossible to reproduce in detail without in-situ measurements. However, to be able to better capture and describe the probable vertical distribution in the source term, three different vertical source layers are implemented. The transition point between the first two source layers is chosen by investigation of the vertical wind profile during this event. A strong variation, especially in the night time, in the main convective wind directions is found at ~200 meters height which is attributed to boundary layer effects. We allow for different amounts of $SO_2$ to be released at three different heights: 0 – 200 and 200 – 1.000 and 1.000 – 4.000 meters. The value of the maximal plume height for the source is an assumption that is supported by the analysis of the 2003 Al-Mishraq fire (Carn et al., 2004) and by comparison with other hot sources, like wild fires (Walter et al.), fissure eruption volcanoes (Beirle et al., 2014) and oil fires (Mather et al., 2007). The choice of using the vertical convective wind profile as the base for the vertical division of the source term is motivated by the source retrieval method explained below.

In conclusion, to obtain the height and time resolved source term outlined above we divide our source term into twelve virtual sources each day, i.e. four time intervals and three vertical layers. Each virtual source has a constant and uniformly distributed release rate. The source estimation problem is now set: given the satellite observations, determine the daily emission rates from the sulphur plant for each virtual source using simulation data from PELLO.

### 4.2. OPTIMIZATION METHOD

Each virtual source, $i$, will give rise to a concentration field that evolves over time. A vertical integration procedure is conducted to render two-dimensional concentration fields, $V_i$, that are commensurable with the satellite data. At any time, the sum of all these fields becomes the collective simulated concentration field, $V_{sim}$, that we compare to the satellite data, $V_{sat}$. Since the problem is linear the fields may be weighted freely, i.e. a change in the release rate for a virtual source equals an identical change in the corresponding concentration field. The problem can therefore be reformulated as finding the optimal set of weights for the fields so that $V_{sim}$ resembles $V_{sat}$ as closely as possible under an appropriate norm. We have chosen the mean square error (MSE) as the norm that describe the degree of agreement between the two fields on a grid. To decrease the influence of spatial translations of the plume (as a result of misrepresentative meteorological data) we use a sparse grid of 1 x 1 degree in the source retrieval method. A robust method is implemented to acquire the best possible linear combination, i.e. weights, of these fields. This method is similar to what has been used in other works (Stohl et al., 2011, Eckhardt et al., 2008, Seibert, 2000). In addition, our method allows for height estimations of the $SO_2$ plume at each position resulting in an improvement of the interpretation of the $SO_2$ load from the satellite data.

We solve the problem by investigating one day at a time in chronological order. Now, the collective simulated field, $V_{sim}$, is obtained by a summation of the known contribution from previous days, $V_0$, and the twelve fields for the current day where a weight vector $\bar{\alpha}$ is introduced, see eq. (1).

$$V_{sim} = V_0 + \sum_{i=1}^{12} \bar{\alpha}(i) V_i \qquad (1)$$

The solution is given by the vector $\bar{\alpha}$ that minimize the difference between $V_{sim}$ and the observed concentration field $V_{sat}$. A complication arise due to the fact that the satellite data, $V_{sat}$, depends on the plume height. In general the plume height has to be observed separately. As no such observations are available we propose that the PELLO simulation results are used to determine the height of the $SO_2$ plume in each position, hence the field $V_{sat}$ is derived using the height information from PELLO. For this reason $V_{sat}$ becomes an implicit function of the $\bar{\alpha}$ vector. A multivariable function based on the derivative-free Nelder-Mead simplex algorithm (Lagarias et al., 1998), was used to find the vector $\bar{\alpha}$ that minimized the MSE of the fields $V_{sat}$ and $V_{sim}$, both dependent of $\bar{\alpha}$. This means that the weights that provide the best numerical fit are found. The optimal value for $\bar{\alpha}$ are solved for under two constraints: first, on each day the total mass of the fields $V_{sim}$ and $V_{sat}$ must agree to within 1%, second $\bar{\alpha}$ must be nonnegative (meaning that negative emissions of $SO_2$ is not acceptable). The source estimation algorithm converges after ~80 iterations in general whereby the vector $\bar{\alpha}$ is obtained.

### 4.3. RESULTS

The method of finding the optimal source term, by means of MSE, given the chosen source design is executed for the entire event. Twelve virtual sources are obtained for each day and are subsequently assembled into one collective source term for the 2016 Al-Mishraq fire. The concentration plumes for both satellite and simulation data are presented in Figure 2 for two different days. The background noise in the satellite data often reaches levels of ~2 DU which is seen in panel a and c in Figure 2.

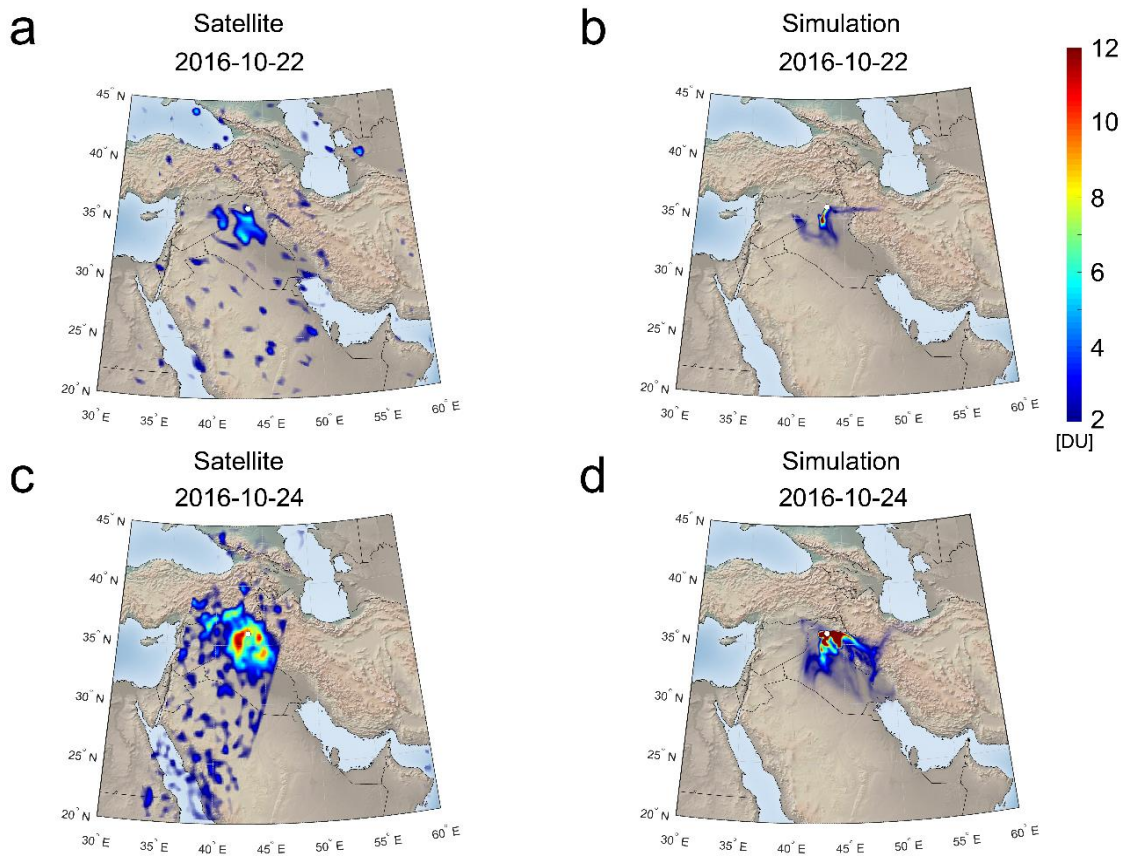

*Figure 2. A spatial comparison between the satellite data of the plume (panels a and c) and the simulated plume (panels b and d). The plumes show the $SO_2$ vertical column densities in Dobson units for October 22 and October 24. Areas with lower concentration than 2.0 Dobson units have been plotted proportionally semi-transperant according to their values.*

The resulting source term, which is the main result of this work, is presented in Table 1 and visualized in Figure 3. A massive fire will give rise to a strong plume rise due to the heat being released which means that the vertical distribution is coupled to the release rate (Briggs, 1969, Briggs, 1971, Briggs, 1972). Indeed, the resulting source term presented in Table 1 shows that on days with a higher total release of $SO_2$ a larger fraction is released on high altitudes compared with days with a smaller total release. Since we are using a dispersion model to infer the $SO_2$ loadings, we may also consider the spatio-temporal agreement between the satellite and the simulated $SO_2$ loadings. Overall the agreement is good, but on some days there is a notable shift between the two fields. In particular on October 21 where the plume observed by the satellite is streching out to the south, while the simulated plume is located more in a southeasterly direction. Likewise on October 25 there is a discrepancy between the satellite data and the simulated concentrations where the satellite derived concentration field is located around 1.5 degree (lat) north of the simulated dito. These discrepancies are attributed to differences between the real weather (giving the observed $SO_2$ plume) and the numerical weather forecast that feeds the dispersion model. When examining the wind field in the weather forecast for October 21 the predominant wind direction locally at Al-Mishraq is north-westerly driving our simulated plume to the south-east. However, the wind field in the entire Mosul area during this day is strongly anisotropic causing small discrepencies to grow with time. The meteorological conditions were difficult to forecast in detail that particular day and this phenomenon aggravates the reproduction of the plume. These spatio-temporal shifts will affect the source estimation method: since the mean square error is minimized under the constraint that the total $SO_2$ loading should be (nearly) equal, the method will compensate for the spatio-temporal shift by bringing additional mass into some parts of the simulated plume (mass that should have been in areas that are unattainable by the simulation result).

*Table 1. The estimated source term for the Al-Mishraq fire 2016. The release rate is given in units of metric tonne per hour and is divided into time intervals of six hours and at three vertical layers. The release rates are assumed to be constant during each time interval and are here presented for each time interval and vertical layer separately. All times are stated in UTC. This data is presented graphically in Figure 3 (where the two upper release heights have been combined). *The satellite data from October 27 has too weak signal to utilize quantatively which means that the source term from October 26 07:00 and forward is difficult to estimate.*

|            | Vertical height | 01:00-07:00 | 07:00-13:00 | 13:00-19:00 | 19:00-01:00 |
|---|---|---|---|---|---|
| October 20 | 0.0 – 0.2 km | 0   | 44   | 46  | 105 |
|            | 0.2 – 1.0 km | 0   | 41   | 56  | 109 |
|            | 1.0 – 4.0 km | 0   | 49   | 53  | 67  |
| October 21 | 0.0 – 0.2 km | 17  | 82   | 86  | 94  |
|            | 0.2 – 1.0 km | 7   | 83   | 85  | 92  |
|            | 1.0 – 4.0 km | 32  | 70   | 79  | 81  |
| October 22 | 0.0 – 0.2 km | 61  | 191  | 200 | 240 |
|            | 0.2 – 1.0 km | 59  | 189  | 241 | 245 |
|            | 1.0 – 4.0 km | 83  | 191  | 212 | 256 |
| October 23 | 0.0 – 0.2 km | 213 | 451  | 450 | 417 |
|            | 0.2 – 1.0 km | 253 | 616  | 589 | 504 |
|            | 1.0 – 4.0 km | 303 | 780  | 803 | 809 |
| October 24 | 0.0 – 0.2 km | 277 | 194  | 239 | 262 |
|            | 0.2 – 1.0 km | 325 | 170  | 241 | 273 |
|            | 1.0 – 4.0 km | 587 | 177  | 230 | 297 |
| October 25 | 0.0 – 0.2 km | 250 | 117  | 144 | 110 |
|            | 0.2 – 1.0 km | 281 | 114  | 135 | 159 |
|            | 1.0 – 4.0 km | 309 | 139  | 162 | 202 |
| October 26 | 0.0 – 0.2 km | 1*  | -*   | -*  | -*  |
|            | 0.2 – 1.0 km | 41* | -*   | -*  | -*  |
|            | 1.0 – 4.0 km | 105*| -*   | -*  | -*  |

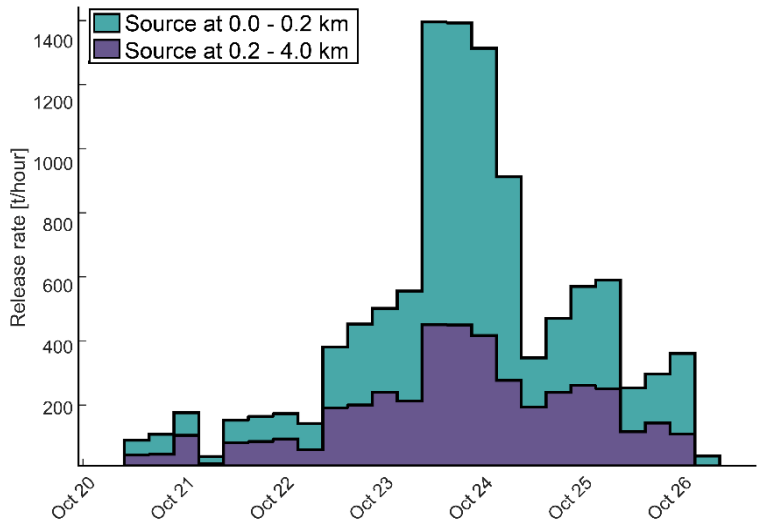

*Figure 3. The source term for the vertical layers (the upper two layers, i.e. 0.2 – 1.0 and 1.0 - 4.0 km have been concatenated) that together constitute the source term. The units are in metric tonnes per hour. This data is numerically presented in Table 1 without the concatenation.*

### 4.4. COMPARISON WITH AIR QUALITY DATA

To validate the source estimation we compare our calculated ground level concentrations against observations from the Turkey National Air Quality Monitoring Network. From satellite data and the dispersion simulations we conclude that the plume passes the southern part of Turkey close to the Iraq border. Of the observation stations in this area Mardin, and Siirt contain clearly increased $SO_2$ levels during the days of interest. In the other stations the background noise is either too high or the observation period too short. In Figure 4 we have compared the calculated ground level concentrations at these two stations and found that simulation data indeed predict an increased concentration. We conclude that our calculated plume centre line probably is located somewhat southeast of the actual plume centre line, most prominent on October 25. Furthermore we can see in our simulation that a small amount of $SO_2$ released in the period October 21 to 22 is transported into Syria and north toward first Mardin on October 23 and later Siirt on 24 east of Mardin. On October 26 the simulated plume passes south of Mardin and Siirt giving rise to the $SO_2$ peaks seen in Figure 4 during October 26-27.

There is a temporal shift in the simulated concentration curves. Looking at the plots in Figure 4 we note that there are early particles arriving at about the right time in both Mardin and Siirt. These are particles that have diffused ahead of the main bulk of the plume. Considering the magnitudes of the peaks, the simulation results are off by approximately an order of magnitude in Mardin but quite on par in Siirt. A good result from a dispersion model should be within a factor 2 of the observed values for 50% of the data points (Chang and Hanna, 2004). In the process of estimating the source term we noted that the simulated plume is located about 1.5 degree south of the satellite plume for October 25-27. To confirm this spatial shift of the simulated results we investigate the concentration at a position 1.5 degrees south of Siirt, which we refer to as "Mock Siirt", and compare the simulated concentration profile with the observed one in (the correctly located) Siirt. The results are shown in Figure 4. Given the uncertainties associated with dispersion modeling, e.g. (Chang and Hanna, 2004), had this been the concentration profile in the real Siirt we would have concluded a very good match between observed and simulated results. Instead, now it seems we have a 1.5 degree shift in the results, which is due to the numerical weather forecast. Note that this shift has very limited influence of the estimation of the total released mass.

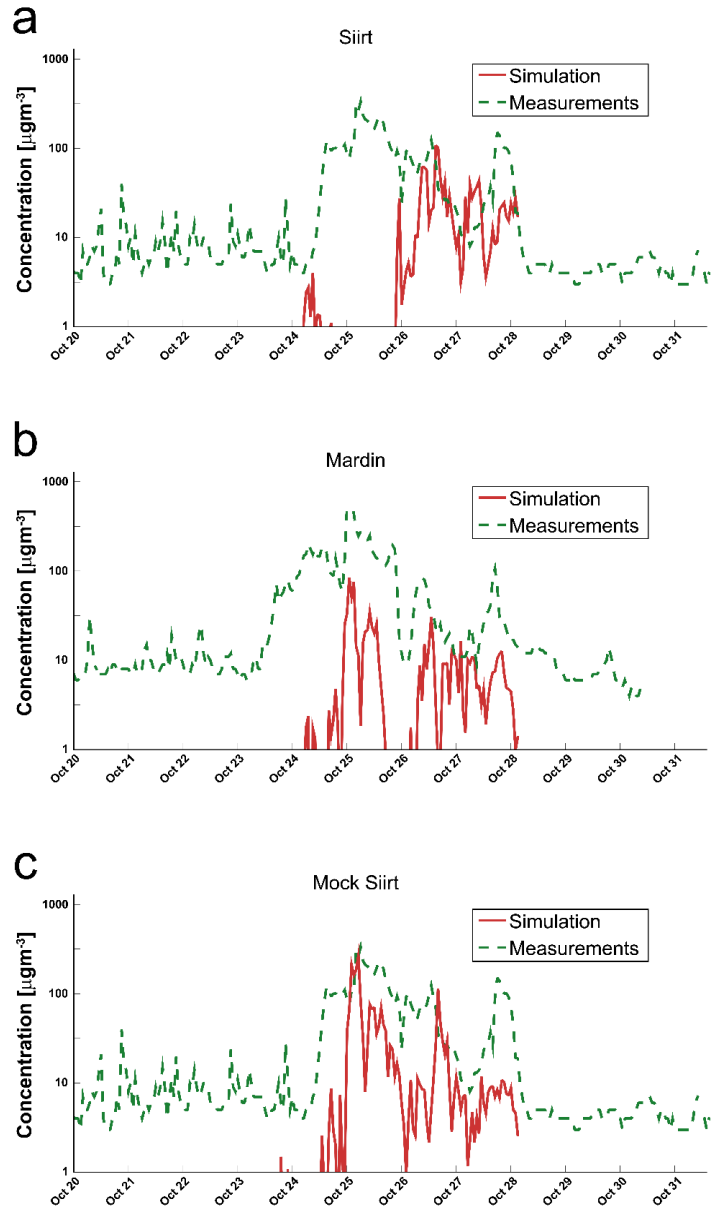

*Figur 4. The sensor data (dashed green) at the two stations Siirt (panel a) and Mardin (panel b) both located in the southeast part of Turkey and the corresponding point concentration from the simulation (solid red). There is a persisting background signal of approximately 10 µgm$^{-3}$ in the detector signal. The simulated data from Mock Siirt (1.5 degrees latitude south of Siirt) compared with measured data results at Siirt is displayed in panel c.*

# 5. RISK AREAS

Exposure to $SO_2$ may cause severe injuries. Common reactions include respiration symptoms as well as irritation, inflammation and burning injuries in the eyes. At acute high concentration the exposure may be lethal. To assess a risk area for the aftermath of the Al-Mishraq fire we have applied the commonly used probit model (Finney, 2009) and utilized time integration of simulation data to obtain the total toxic load for each day separately during the period of interest for ground-level concentrations. Two different parameter sets have been utilized for this analysis. The first corresponds to the frequently used AEGL values (Green-Book, 1992, National Research Council, 2000) and describes the risk area, panel a. The second is based directly on published human data on casualty assessments (Frank et al., 1962, Cohen et al., 1973, Amdur et al., 1953) and excludes any safety margins, panel b. Given a toxic load-field we identify the region where at least 5% of a static population reaches level 1 (green) and level 2 (red) for the two parameter sets.

The general definition of AEGL-1 (National Research Council, 2012) is

*"the airborne concentration (expressed as parts per million or milligrams per cubic meter [ppm or mg/m3]) of a substance above which it is predicted that the general population, including susceptible individuals, could experience notable discomfort, irritation, or certain asymptomatic, nonsensory effects. However, the effects are not disabling and are transient and reversible upon cessation of exposure."*

and for AEGL-2 (National Research Council, 2012) is

*"the airborne concentration (expressed as ppm or mg/m3) of a substance above which it is predicted that the general population, including susceptible individuals, could experience irreversible or other serious, long-lasting adverse health effects or an impaired ability to escape."*.

Note that for $SO_2$ specifically, the AEGL-2 values refer to asthmatic or otherwise sensitive individuals (National Research Council, 2010).

For the casualty assessments, level 1 refers to light injuries and level 2 refers to severe injuries for a general healthy population which is also the nomenclature used in Figure 5b. The unions of the daily regions were collected into final areas for each parameter set, see Figure 5. The red area, indicating severe injuries, in panel b is in subgrid size and is therefore not visible.

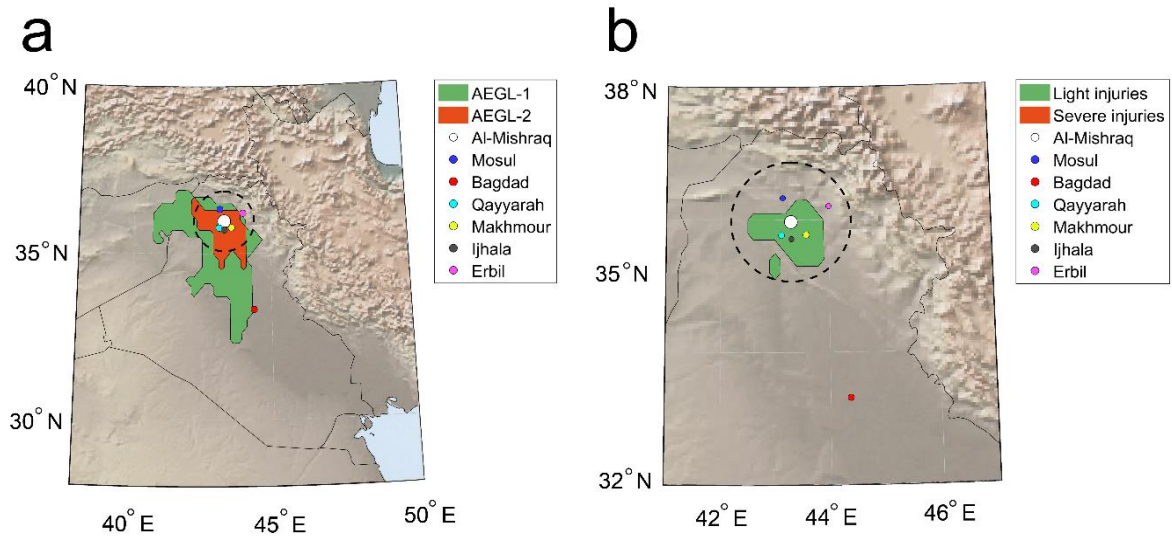

*Figure 5. Risk and casualty area assessments yield spatial distributions that are to be interpreted as the regions wherein there was a realistic probability to reach level 1 (green area), or level 2 (red area) on any day during the Al-Mishraq fire. Panel a shows the risk area according to AEGL-data and panel b shows the casualty assessment area. The circle indicates a 100 km distance from Al-Mishraq.*

The United Nation Environment Programme (UNEP) reports that the Directorate of Health and the World Health Organization treated over 1000 persons for suffocation symptoms in the Iraqi cities Qayyarah, Makhmour and Ijhala (Newsdesk, 2016). These three cities are depicted in Figure 5 and they all reside south of Al-Mishraq. The parameter sets give rise to completely different areas where the AEGL risk areas (panel a) are extensive while the areas obtained for the casualty assessments (panel b) predicts the actual injury outcome very accurately.

# 6. DISCUSSION

It has previously been shown that satellite measurements of sulphur dioxide provide a tool for assessing the release rate of $SO_2$ from large events like volcanic eruptions (Theys et al., 2013) to smaller events like the Al-Mishraq fire in 2003 (Carn et al., 2004). What we show here is that with GOME-2 data being readily available, even in near real-time, 1) a rapid source term estimation can be made to determine the release rate of $SO_2$ from an industrial accident, in this case caused by antagonists, to 2) be used for forecasting of $SO_2$ concentration in the region by use of dispersion models, and 3) to, by using the dispersion model results, estimate in which areas the dispersed $SO_2$ constitutes a health risk. In lieu of in situ $SO_2$ detectors providing the ground truth, the information derived using the method outlined in this paper may prove very useful to health care officers or military commanders. While satellite images show the extent of the $SO_2$ plume (given no interference from clouds) the drawback is that this information is only available at the discrete time slots determined by the satellite passage times over the area and secondly that $SO_2$ concentrations are typically presented as total columns, i.e. the concentration is integrated over all heights in each vertical column. The advantage of using satellite informed dispersion models is that they complete the picture by filling in the gaps between consecutive satellite observations (which for MetOp-A and MetOp-B is once every 24 hours in the present case) as well as providing height-resolved concentrations of $SO_2$. The latter is crucial for estimating the risk to the population, and in this case deployed military troops, as they are only vulnerable to $SO_2$ at ground level.

Based on the fractioned satellite data we have presented an estimate of the source term for the 2016 Al-Mishraq fire. The source term is time-dependent and also divided into three vertical layers to address the anisotropic nature of the atmospheric transport conditions. The total amount of released $SO_2$ has been estimated to 92 kt for the first 6 days of the fire. The release rate of the last day of the fire, October 26, is difficult to estimate due to weak detector signal. The SEVIRI images do show that some $SO_2$ is released on this day, which the source estimate does not include. The average release rate is 15 kt/day which can be compared with the average release rate of the fire in 2003 of 21 kt/day reported in (Carn et al., 2004) (although that source term contains ten days of no release, hence a more accurate average would be close to 35 kt/day). The 2016 fire was less extensive, with both average and maximum emission rates lower than those estimated for the 2003 fire. Given the recent history in Iraq it is likely that the stockpiles of sulphur were reduced since 2003.

The source term has been validated by using it for dispersion model runs that are compared against independent $SO_2$ measurements from the sites Mardin and Siirtin in the Turkey National Air Quality Monitoring Network. The comparison shows a reasonable agreement between model results and measurements. When making a sensitivity analysis we find that, for the measurements in Turkey, a very good fit, both in terms of arrival time and magnitude, would have been achieved had the $SO_2$ plume been shifted north by about 1.5 degrees (lat). Although this sensitivity is undesired, we should bear in mind that there are large uncertainties involved in both the source estimation and dispersion modeling: the target uncertainty associated with GOME-2 $SO_2$ measurements is 50% (Hassinen et al., 2016) and the satellites only pass once per day (and nearly at the same time 07:00), the dispersion model is a simplification of the processes in the atmosphere and there is an uncertainty associated with the numerical weather forecasts employed to drive the dispersion model. As it is difficult to tell whether these errors cancel or reinforce each other it is hard to estimate the overall uncertainty associated with the source estimate. In (Stohl et al., 2011) it was suggested that this uncertainty is 50%. Given the target uncertainty of GOME-2, we believe that a conservative estimate of the source estimation uncertainty should be larger.

We remark that the source term obtained by this method is indeed the optimal solution to the source estimation problem given the chosen source design (constant release rate in 6 hour time intervals and at three vertical layers) using MSE as the norm. There are though large uncertainties involved in the system. The inherent uncertainties of the satellite data is enhanced in the presence of clouds (Valks, 2015) which is the case for the last days during this event. The impact of the clouds is to some extent unknown but depends on the relative heights of the clouds and the plume. Furthermore, dispersion modeling rely heavily on meteorological data. Errors in the meteorological data will propagate through the model and cause errors in the model outcome. In lack of perfect meteorological data there is no source term that will give rise to an exact match between simulated and measured concentration fields. During this event we notice that the meteorological data predicts most of the convective winds correctly. However, the plume drift to the north during the last days is captured but to a lesser extent than is seen from satellite data. We should bear in mind that the uncertainties in $SO_2$ measurements and meteorology data propagates to the estimated source term which is to be regard as a good estimate of the true source term.

The risk areas associated with the Al-Mishraq fire were presented to show how satellite measurement of $SO_2$ can feed dispersion models and effect models to aid health care officers, or military commanders, in making informed decisions of how to act in situations where in situ measurements are unavailable. Given that $SO_2$-related injuries were reported from health centres in Qayyarah, Makhmour, and Ijhala and, to our knowledge, nowhere else and that these health centres lie well within the casualty assessment area presented constitutes another form of validation for our method and source term.

**Acknowledgements**: We thank The Swedish Ministry of Defence for the financial support that made this work possible. We are grateful to Martin Raspaud, SMHI, for providing us with Meteosat-10 images over Iraq. We also thank Seppo Hassinen, FMI, and Pascal Hedelt, DLR, for guiding us to the GOME-2 data.